\documentclass[referee,useAMS]{mn2e}

\usepackage{graphicx}
\usepackage{amsmath}
\font\twelvei = cmmi10 scaled\magstep1
       \font\teni = cmmi10 
\font\mbf = cmmib10 scaled\magstep1
       \font\mbfs = cmmib10 \font\mbfss = cmmib10 scaled 833
\font\msybf = cmbsy10 scaled\magstep1
       \font\msybfs = cmbsy10 \font\msybfss = cmbsy10 scaled 833
\def\lsim{\raise0.3ex\hbox{$<$}\kern-0.75em{\lower0.65ex\hbox{$\sim$}}}
\def\gsim{\raise0.3ex\hbox{$>$}\kern-0.75em{\lower0.65ex\hbox{$\sim$}}}

\title[QPOs in Radiative Transonic Flow]
{Quasi Periodic Oscillations in a Radiative Transonic Flow: Results of a Coupled Monte Carlo-TVD Simulation}
\author[Sudip K. Garain, Himadri Ghosh, Sandip K. Chakrabarti]
{Sudip K. Garain\thanks{sudip@bose.res.in}$^{1}$, Himadri Ghosh\thanks{ehimadri@gmail.com}$^{2}$, Sandip K. Chakrabarti\thanks{chakraba@bose.res.in}$^{1,2}$\\
$^{1}$S. N. Bose National Centre for Basic Sciences, Salt Lake,
              Kolkata 700098, India\\
$^{2}$Indian Centre for Space Physics, Chalantika 43, Garia Station Rd.,
             Kolkata, 700084, India}
\begin{document}
\date{}

\maketitle

\label{firstpage}

\begin{abstract}
Low and intermediate frequency quasi-periodic oscillations (QPOs) in black hole candidates are believed to be
due to oscillations of the Comptonizing regions in an accretion flow. Assuming that the general 
structure of an accretion disk is a Two Component Advective Flow (TCAF), we 
numerically simulate the light curves emitted from an accretion disk for different 
accretion rates and find how the QPO frequencies vary. We use a standard 
Keplerian disk residing at the equatorial plane as a source of soft photons. These soft 
photons, after suffering multiple scattering with the hot electrons of the 
low angular momentum, sub-Keplerian, flow emerge out as hard radiation. 
The hydrodynamic and thermal properties of the electron cloud is simulated 
using a Total Variation Diminishing (TVD) code. The TVD code is then 
coupled with a radiative transfer code which simulates the 
energy exchange between the electron and radiation using Monte Carlo 
technique. The resulting localized heating and cooling are included also.
We find that the QPO frequency increases and the spectrum becomes 
softer as we increase the Keplerian disk rate. However, the spectrum becomes harder
if we increase the sub-Keplerian accretion rate. We 
find that an earlier prediction that QPOs occur when the infall 
time scale roughly matches with the cooling time scale,
originally obtained using a power-law cooling, remains valid even for Compton cooling.  
Our findings agree with the general observations of low frequency QPOs in black hole
candidates.

\end{abstract}


\section{Introduction}
Quasi-periodic oscillations (QPOs) observed in X-rays are very important features for the study
of accreting black holes. 
Observations and possible explanations of QPOs in black hole candidates have been reported quite 
extensively in the literature (e.g., Chakrabarti et al. 2008a and the references therein). 
They are believed to be the manifestations of some regular time-dependent properties of the 
underlying accretion flows and happen to be closely connected to the spectral state transitions (Remillard
\& McClintock 2006).  
X-ray transient sources in our galaxy exhibit various types of QPOs with 
frequencies ranging from $ ~ 0.001 - 0.01$ Hz to a few hundreds
of Hz (Morgan, Remillard \& Greiner 1997; Paul et al. 1998; Yadav et al. 1999; Remillard \& McClintock 2006). 
However, the quasi-periodic behavior is not always observed. More common feature is
the erratic variation of photon count rates even in a timescale of seconds.

It has also been reported that there should exist a correlation between the 
QPO frequency and the spectral index (Remillard \& McClintock 2006, Chakrabarti et al. 2008b; 
Chakrabarti, Dutta \& Pal 2009; Debnath, Chakrabarti \& Nandi 2010; Stiele et al. 2013). 
These authors reported a rise in photon index with increasing centroid frequency of QPOs
for various black hole candidates
during the onset of the outbursts. It is also reported that during the decay of the
outburst, the trend is opposite. The correlation does not 
depend much on any individual outburst, rather they follow a general trend. 

The spectral properties of the black hole candidates can be explained using 
several models, which generally include two components, namely, a Keplerian disk
and a hot `corona', only the nature of the corona varies from model to model. Sunyaev and Titarchuk (1980, 1985) pointed out that any model having 
a hot electron cloud should be enough to produce a power-law component which is 
observed especially in the hard state. Judging from the rapidity in which the 
spectral properties change, a realistic model would be where the `corona' 
itself moves at a short timescale. The so-called two-component advective flow (TCAF) solution 
(Chakrabarti \& Titarchuk 1995,
hereafter CT95) has a standard Shakura-Sunyaev Keplerian disk 
(Shakura \& Sunyaev 1973, hereafter SS73) on the equatorial plane and a sub-Keplerian 
(low angular momentum) accreting halo on the top and bottom of the Keplerian disk.
 Thus, both the 
flows are moving, albeit in different time scales. It has been recently shown (Giri \& Chakrabarti 2013)
through numerical simulation of a radiative viscous flow, that such a TCAF configuration, 
envisaged in CT95, is not only achievable, but is stable as well. 
The sub-Keplerian component (halo) produces a centrifugal barrier dominated hot region
in the inner disk, which may or may not develop into a shock (Chakrabarti 1989). This region, including the 
post-shock region, where the centrifugal force dominates,
behaves like the Compton cloud of hot electrons, while the Keplerian disk produces 
the soft photons. The high energy, power-law component of the spectrum is 
due to inverse Comptonization of the soft photons by hot electrons supplied
by the sub-Keplerian flow (CT95; Ghosh, Chakrabarti \& Laurent 2009).
In this model, the oscillation of X-ray intensity is caused by the
the oscillation of the post-shock (Comptonizing) region 
(Molteni, Sponholz \& Chakrabarti 1996, hereafter MSC96; 
Chakrabarti \& Manickam 2000, hereafter CM00) which reprocesses different amounts of 
soft photons at different phases of oscillation. 
The numerical simulation of low-angular momentum accretion flows
including the thermal cooling (MSC96; Chakrabarti, Acharyya \& Molteni 2004, hereafter CAM04) 
or dynamical cooling (through outflows, e.g., Ryu, Chakrabarti \& Molteni 1997) 
show clearly that the shocks oscillate with frequencies similar to the observed QPO frequencies.

In case the accretion rates are not constant over a period, but evolve,
as is observed in outburst sources, the shocks propagate, mainly due to increased cooling
of the post-shock region which reduces the pressure.
The typical evolution of the QPOs during the outburst and the decaying phases is
explained using the propagating oscillatory shock (POS) model (Chakrabarti et al. 
2005; Chakrabarti  et al. 2008a; Debnath et al. 2008, 2010). 
In this model, the sudden enhancement of viscosity causes the Keplerian disk 
accretion rate ($\dot{m}_d$) to increase with time and 
the shock advances towards the black hole due to 
increasing cooling in the post-shock region. This
increases the QPO frequency as the spectrum softens. After withdrawal 
of viscous effects, $\dot{m}_d$ starts to decrease, and the shock recedes. 
This decreases 
the QPO frequency while the spectrum hardens due to dominance of the Compton cloud,
which becomes difficult to cool in presence of lower $\dot{m}_d$. Thus, generally shock oscillation
model is capable of explaining most of the observations relating to low and intermediate frequency QPOs.

So far, while the shock-oscillation was generally proposed to be the cause of QPOs, and the numerical
simulations generally showed that this could very well be the case (MSC96; CAM04), 
there was no attempt to include the Keplerian disk into the simulation 
and only enhanced power-law cooling was used as a proxy to inverse Comptonization. 
In the present paper, we use a time dependent coupled hydrodynamic and 
radiative transfer code (Ghosh et al. 2011; Garain, Ghosh \& Chakrabarti 2012) which includes 
both the source of seed photons
and actual computation of the production of hard photons through Comptonization process.
We analyze the resulting light curves to find the QPOs and other timing 
properties using the NASA HEAsoft package, as though our outgoing photons correspond to the observed 
photons. We clearly find a correlation between the flow parameters and the 
QPO frequencies. We vary the  accretion rates and study the dependence.  
In the next Section, we discuss our simulation set up and the procedure 
to compute the hydrodynamic and the radiative processes inside the 
accretion flow. In \S 3, we present the results of our simulations. 
Finally, in \S 4, we make concluding remarks.

\section{Simulation procedure}

\begin{figure}
\begin{center}
\includegraphics[width=8cm,height=6cm]{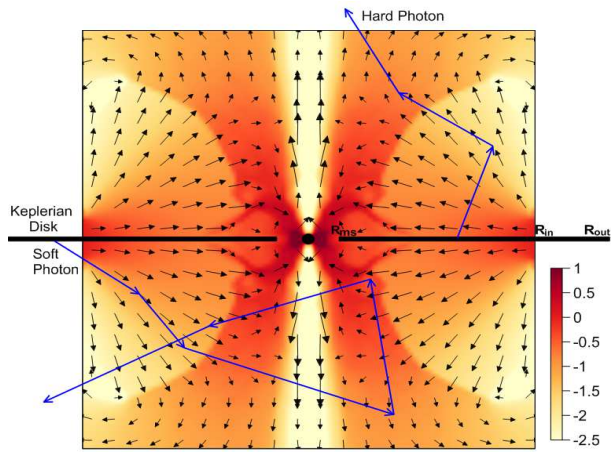}
\caption{The schematic diagram of our simulation set up. The velocity vectors
of the infalling matter are shown. The colors show the normalized density 
in a logarithmic scale. The zig-zag trajectories (blue online) are the 
typical paths followed by the photons. (A color version is available online.)
}
\end{center}
\end{figure}

In Figure 1, we present a schematic diagram of our simulation set up. 
We consider a two-component advective flow (TCAF) as our initial configuration. 
Here, a standard Keplerian disk is sandwiched between a low angular momentum, 
sub-Keplerian flow located above and below. The soft photons emerging out 
of the Keplerian disk are intercepted and reprocessed via Compton or 
inverse-Compton scattering by the sub-Keplerian matter, both in the post-shock 
and the pre-shock regions. For the sake of concreteness, we consider the sub-Keplerian 
flow with a specific angular momentum $\lambda=1.73$. The sub-Keplerian matter 
is injected from the outer boundary at $R_{in} = 100 r_g$ ($r_g=2GM_{bh}/c^2$, where 
$G$, $M_{bh}$ and $c$ are the gravitational constant, mass of the black hole 
and the velocity of light respectively). 
The outer edge of the Keplerian disk is located at $R_{out} = 200 r_g$ 
and it extends inside up to the marginally stable orbit $R_{ms} = 3 r_g$. 
At the center, a black hole of mass $M_{bh}$ is located. 


\subsection{Sub-Keplerian flow}
A realistic accretion disk is three-dimensional. Assuming 
axisymmetry, and thus reducing the problem to two dimensions, we simulate 
the flow dynamics using a finite difference 
method which uses the principle of total variation diminishing (TVD) to 
carry out hydrodynamic simulations (see, Ryu et al. 1997 and references 
therein; Giri et al. 2010). To model the initial injection of matter, 
we consider an axisymmetric flow of gas in the pseudo-Newtonian gravitational 
field of a black hole of mass $M_{bh}$, located at the center in the cylindrical 
coordinates $[R,\theta,z]$. 
We assume that the gravitational field of the black hole can be described by Paczy\'{n}ski 
\& Wiita (1980) pseudo-Newtonian potential,
$$
\phi(r) = -{GM_{bh}\over(r-r_g)}, 
$$
where, $r=\sqrt{R^2+z^2}$. 
We assume a polytropic equation of state for the accreting and outflowing
matter, $P=K \rho^{\gamma}$, where, $P$ and $\rho$ are the isotropic pressure 
and the matter density respectively, $\gamma$ is the adiabatic index (assumed
to be constant throughout the flow, and is related to the polytropic index 
$n$ by $\gamma = 1 + 1/n$) and $K$ is related to the specific entropy of 
the flow $s$. $K$ is not constant but is allowed to vary due to radiative 
processes. The details of the hydrodynamics part of the code is described in 
Molteni, Ryu \& Chakrabarti (1996) and Giri et al. (2010).

Our computational box occupies one quadrant of the $R-z$ plane with 
$0 \leq R \leq 100$ and $0 \leq z \leq 100$. The incoming gas enters 
the box through the outer boundary, located at $R_{in} = 100$. We 
have chosen the density of the incoming gas ${\rho}_{in} = 1$ for 
convenience. In the absence of self-gravity, the density 
is scaled out, rendering the simulation results valid for any accretion rate. 
We put the sound speed $a$ (i.e., temperature) of the flow and the incoming 
velocity at the outer boundary in a way which makes the specific energy as that of the injected energy. 
In order to mimic the horizon of the black hole at the Schwarzschild radius, 
we place an absorbing inner boundary at $r = 1.5 r_g$, inside which all 
material is completely absorbed into the black hole. For the background matter 
(required to avoid division by zero), we use a stationary gas with density 
${\rho}_{bg} = 10^{-6}$ filling up the entire grid. The initial sound speed 
(or, temperature) in the whole grid is chosen to be the same as that of the incoming gas. 
Hence the incoming matter has a pressure $10^6$ times larger than that 
of the background matter.  Thus we anticipate that in about one dynamical time, 
the initial grid matter will be totally replaced by the incoming 
low angular momentum matter. 
All the calculations were performed with $512 \times 512$ equi-spaced cells. Thus, each 
grid has a size of $0.19$ in units of the Schwarzschild radius. 

\subsection{Keplerian disk}

The Keplerian disk is assumed to be optically thick and the opacity due 
to free-free absorption is more important than the opacity due to scattering. 
The soft photons are produced from the Keplerian disk, the inner edge of 
which has been kept fixed at the marginally stable orbit $R_{ms}$, while 
the outer edge is located at $R_{out}$ ($200 r_g$). The source of the soft 
photon is a multicolor blackbody spectrum coming from a standard 
(SS73) disk. The emission is blackbody type with the local surface temperature 
(SS73):

\begin{equation}
T(r) \approx 5.48 \times 10^7 (M_{bh})^{-1/4}(\dot{m}_d)^{1/4} (2r)^{-3/4}
 \left[1- \sqrt{\frac{3}{r}}\right]^{1/4} \mathrm{~K}.
\end{equation}

Photons are emitted from both the top and the bottom surfaces of the disk at
each radius. The total number of photons emitted from the disk surface at
radius $r$ is obtained by integrating over all frequencies ($\nu$) and
is given by,
\begin{equation}
n_\gamma(r) = \frac{4\pi}{c^2} \left(\frac{kT}{h}\right)^3 \times 1.202
\mathrm{~cm^{-2}~s^{-1}.}
\end{equation}
Thus, the disk between radius $r$ to $r+\delta r$ produces $dN(r)$ number
of soft photons:
\begin{equation}
dN(r) =  4\pi r\delta r n_{\gamma}(r) \mathrm{~s^{-1}}.
\end{equation}

In the Monte-Carlo simulation, we incorporate the directional effects 
of photons coming out of the Keplerian disk with the maximum number of 
photons emitted in the $z$-direction and minimum number of photons are 
generated along the plane of the disk. Thus, in the absence of photon 
bending effects, the disk is invisible as seen edge on. The position of 
each emerging photon is randomized using the distribution function (Eq. 3). 
In the above equations, the mass of the black hole $M_{bh}$ is measured 
in units of the mass of the Sun ($M_\odot$), the disk accretion rate 
$\dot{m}_d$ is in units of mass Eddington rate.
We chose $M_{bh} = 10$ in the rest of the paper. 

\subsection{Radiative process}

We simulate the Compton scattering between the soft radiation from the 
Keplerian disk and the hot electrons in the sub-Keplerian flow using a 
Monte-Carlo code.
To begin a Monte-Carlo simulation, we generate photons from the Keplerian disk with
randomized locations as mentioned in the earlier section. The energy of the 
soft photons at radiation temperature $T(r)$ is calculated using the Planck's 
distribution formula, where the number density of the photons ($n_\gamma(E)$) 
having an energy $E$ is expressed by (Pozdnyakov, Sobol \& Sunyaev 1983),
\begin{equation}
n_\gamma(E) = \frac{1}{2 \zeta(3)} b^{3} E^{2}(e^{bE} -1 )^{-1}, 
\end{equation}
where, $b = 1/kT(r)$ and $\zeta(3) = \sum^\infty_1{l}^{-3} = 1.202$, the 
Riemann zeta function. Using another set of random numbers, we obtain the 
direction of the injected photon and with yet another random number, we 
obtain a target optical depth $\tau_c$ at which the scattering takes place. 
The photon is followed within the sub-Keplerian matter till the optical depth ($\tau$)
reached $\tau_c$. The increase in optical depth ($d\tau$) during its traveling of
a path of length $dl$ inside the sub-Keplerian matter is given by: $d\tau = \rho_n \sigma dl$, 
where $\rho_n$ is the electron number density.

The total scattering cross section $\sigma$ is given by Klein-Nishina formula:
\begin{equation}
\sigma = \frac{2\pi r_{e}^{2}}{x}\left[ \left( 1 - \frac{4}{x} - 
\frac{8}{x^2} \right) ln\left( 1 + x \right) + \frac{1}{2} + \frac{8}{x} - 
\frac{1}{2\left( 1 + x \right)^2} \right],
\end{equation}
where, $x$ is given by,
\begin{equation}
x = \frac{2E}{m c^2} \gamma \left(1 - \mu \frac{v}{c} \right),
\end{equation}
$r_{e} = e^2/mc^2$ is the classical electron radius and $m$ is the mass of the electron.

We have assumed here that a photon of energy $E$ and momentum $\frac{E}{c}\boldsymbol{\Omega}$
is scattered by an electron of energy $\gamma mc^{2}$ and momentum 
${\boldsymbol{p}} = \gamma m \boldsymbol{v}$, 
with $\gamma = \left( 1 - \frac{v^2}{c^2}\right)^{-1/2}$ and $\mu = \boldsymbol{\Omega.v}$.
At this point, a scattering is allowed to take place. The photon selects 
an electron and the energy exchange is computed using the Compton or 
inverse-Compton scattering formula. The electrons are assumed to obey relativistic 
Maxwell distribution inside the sub-Keplerian matter.  The number $dN(\boldsymbol{p})$ of 
Maxwellian electrons having momentum between $\boldsymbol{p}$ to $\boldsymbol{p} + d\boldsymbol{p}$ 
is expressed by,
\begin{equation}
dN(\boldsymbol{p}) \propto exp[-(p^2c^2 + m^2c^4)^{1/2}/kT_e]d\boldsymbol{p}.
\end{equation}

\subsubsection{Calculation of energy exchange using Monte-Carlo code:}

We divide the Keplerian disk in different annuli of width $D(r)=0.5 r_g$.
Each annulus having mean radius $r$ is characterized by its average 
temperature $T(r)$. The total number of photons emitted from the disk 
surface of each annulus can be calculated using Eq. 3. 
This total number comes out to be $\sim~10^{46}$ per second for $\dot{m}_d = 0.1$ Eddington rate.
In reality, one cannot inject these many number of photons in a Monte-Carlo simulation
because of the limitation of computational time. So we replace this large number of photons
by a lower number of bundles of photons, say, $N_{comp}(r)~\sim~10^7$ 
and calculate a weightage factor,
$$
f_W = \frac{dN(r)}{N_{comp}(r)}.
$$
Clearly, from annulus to annulus, the number of photons in a bundle will vary. 
This is computed from the standard disk model and is used to compute the change of energy 
by Comptonization. When this injected photon is inverse-Comptonized
(or, Comptonized) by an electron in a volume element of size $dV$, 
we assume that $f_W$ number of photons have suffered a similar
scattering with the electrons inside the volume element $dV$. If the energy
loss (gain) per electron in this scattering is $\Delta E$, we multiply this
amount by $f_W$ and distribute this loss (gain) among all the electrons inside
that particular volume element. This is continued for all the $10^7$ bundles of photons
and the revised energy distribution is obtained. In this way, the injected 
energy in the sub-Keplerian flow is divided into matter and photon unevenly 
depending on the optical depth of local matter as seen from the Keplerian disk.

\subsubsection{Computation of the temperature distribution after cooling} 

Since the hydrogen plasma considered here is ultra-relativistic ($\gamma=\frac{4}{3}$ 
throughout the hydrodynamic simulation), thermal energy per particle is 
$3k_BT_e$, where, $k_B$ is Boltzmann constant, $T_e$ is the temperature of electron.
The electrons are cooled by the inverse-Comptonization of the soft photons
emitted from the Keplerian disk. The protons are cooled because of the
Coulomb coupling with the electrons. Total number of electrons inside
any box with the center at grid location $(ir,iz)$ is given by,
\begin{equation}
dN_e(ir,iz) = 4\pi rn_e(ir,iz)dRdz, 
\end{equation}
where, $n_e(ir,iz)$ is the electron number density at $(ir,iz)$ grid, and
$dR$ and $dz$ represent the grid size along $R$ and $z$ directions respectively. So
the total thermal energy in any box is given by, $$3k_BT_e(ir,iz)dN_e(ir,iz) = 12\pi
rk_BT_e(ir,iz)n_e(ir,iz)dRdz,$$ where, $T_e(ir,iz)$ is the temperature at $(ir,iz)$
grid. We calculate the total energy loss (gain) $\Delta E$ of electrons inside the
box according to what is presented above and subtract that amount to get the
new temperature of the electrons after each scattering inside that box as
\begin{equation}
k_BT_{e,new}(ir,iz) = k_BT_{e,old}(ir,iz)-\frac{\Delta E}{3dN_e(ir,iz)}.
\end{equation}

\subsection{Coupling of hydrodynamic and radiative codes}

The hydrodynamic and the radiative transfer codes are coupled together following 
a similar procedure as in Ghosh et al. (2011) and
Garain et al. (2012), and this will not be repeated here in detail. However, for
the sake of completeness, we outline the steps here.
Once a quasi steady state is achieved using the non-radiative 
hydro-code, we compute the radiation spectrum using the
Monte-Carlo code in the same way mentioned above. This is the first approximation 
of the spectrum. To include the cooling in the coupled code, we follow these steps: 
(i) we calculate the velocity, density and temperature profiles of the
electron cloud from the output of the hydro-code. 
(ii) Using the Monte-Carlo code we calculate the spectrum.
(iii) Electrons are cooled (heated up) by the inverse-Compton (Compton) scattering. 
We calculate the amount of heat loss (gain) by the electrons and its new temperature 
and energy distributions and
(iv) taking the new temperature and energy profiles as initial condition, 
we run the hydro-code for some time. Subsequently, we repeat the 
steps (i-iv). In this way, we see how the spectrum is modified as the time proceeds.

All the simulations are carried out assuming a stellar mass black hole 
$(M_{bh} = 10{M_\odot})$.
The procedures remain equally valid for massive/super-massive black holes. 
However, for massive black holes we also need to include bremsstrahlung 
(both thermal and magnetic)
processes in the Monte-Carlo simulation for the supply for seed photons.

\section{Results and Discussions}
For a particular simulation, we use the Keplerian disk rate ($\dot{m}_d$) and the
sub-Keplerian halo rate ($\dot{m}_h$) as parameters. The specific energy ($\epsilon$) 
and the specific
angular momentum ($\lambda$) determine the hydrodynamics (shock location, number 
density and velocity variations etc.) 
and the thermal properties of the sub-Keplerian matter.

{\tiny
\begin {tabular}[h]{|c|c|c|c|c|c|c|c|c|c|c|c|}
\hline
\multicolumn{12}{|c|}{Table 1: Parameters used for the simulations and a summary of results.}\\
\hline
ID & $\dot{m}_d$ & $\dot{m}_h$ & ${\rm <R_{sh}>}$ & $\nu_{QPO}$ & $<\alpha>$ & ${\rm E_{th}}$ & ${\rm \dot{E}}$ & $t_{in}\over t_{cool}$ & ${\rm E_{bb}}$ & ${\rm E_{sc}}$ & Q \\
\hline
C1  & 1e-4 & 0.1  & 25.75 &   -   & 0.826 & 2.17e33  & 1.93e34  & 0.694 & 8.21e32 & 1.47e34 & 17.97\\
C2  & 2e-4 & 0.1  & 22.82 & 10.63 & 0.853 & 1.69e33  & 1.82e34  & 0.844 & 1.35e33 & 1.87e34 & 13.87\\
C3  & 3e-4 & 0.1  & 20.39 & 12.34 & 0.868 & 1.52e33  & 2.90e34  & 0.954 & 1.98e33 & 2.72e34 & 13.71\\
C4  & 4e-4 & 0.1  & 18.62 & 14.63 & 0.873 & 1.35e33  & 3.20e34  & 0.944 & 2.37e33 & 3.26e34 & 13.79\\
C5  & 5e-4 & 0.1  & 18.33 & 22.74 & 0.901 & 1.13e33  & 3.85e34  & 1.071 & 2.66e33 & 2.90e34 & 10.88\\
C6  & 1e-3 & 0.1  & 15.02 &  18.2 & 1.074 & 7.40e32  & 2.86e34  & 0.993 & 4.78e33 & 3.73e34 & 7.81\\
C7  & 1e-2 & 0.1  &  3.4  &   -   & 1.139 & 5.51e32  & 1.69e35  & 3.558 & 4.53e34 & 2.11e35 & 4.7 \\
C8  & 1e-1 & 0.1  &  3.6  &   -   & 1.102 & 3.79e32  & 1.04e36  & 33.24 & 4.34e35 & 1.65e36 & 3.8\\
C9  & 3e-4 & 0.05 & 23.53 & 10.74 & 0.998 & 9.63e32  & 1.11e34  & 0.859 & 1.24e33 & 1.13e34 & 9.12\\
C10 & 3e-4 & 0.15 & 18.37 & 22.80 & 0.797 & 1.60e33  & 4.22e34  & 1.031 & 2.48e33 & 5.26e34 & 21.25\\
C11 & 3e-4 & 0.2  & 16.78 &   -   & 0.757 & 1.91e33  & 7.56e34  & 1.049 & 2.93e33 & 7.12e34 & 24.26\\
\hline
\end{tabular}\\
}

In Table 1, we show the parameters used for the simulations and a summary
of the results. For all the cases, the specific energy $\epsilon = 0.0021$
and the specific angular momentum $\lambda = 1.73$ of the sub-Keplerian flow
at the outer boundary have been used. The energy and the angular momentum are measured
in the unit of $c^2$ and $r_gc$, respectively. 
The case IDs are given in the first Column. Columns 2 and 3 show the
Keplerian disk rate ($\dot{m}_d$) and sub-Keplerian halo rate ($\dot{m}_h$),
respectively. Both are in units of the Eddington rate. In Column 4,
we present the time averaged shock location in $r_g$ near the equatorial plane.
For some combinations of $\dot{m}_d$ and $\dot{m}_h$, we find QPOs, about which we shall comment later.
In Column 5, we list the QPO frequencies in Hz. The time averaged spectral
slope ($\alpha,~I(E)\propto E^{-\alpha}$) is given in Column 6. The thermal energy content
(${\rm E_{th}}$ in erg) and the cooling rate (${\rm \dot{E}}$ in erg/sec) in the post-shock
region only are given in the Column 7 and 8, respectively. In Column 9,
we show the ratio of the infall time scale ($t_{in}$) and the cooling
time scale ($t_{cool}$). $t_{cool}$ is computed by taking the ratio of
${\rm E_{th}}$ and ${\rm \dot{E}}$. Column 10 gives the total initial
energy per second (${\rm E_{bb}}$ in erg/sec) of all those very photons 
which emit from the Keplerian disk and
suffer at least one scattering inside the Compton cloud before leaving
the system. The total final energy per second (${\rm E_{sc}}$ in erg/sec) 
of these particular photons, when
they leave the system, is listed in Column 11. The enhancement factor (Q),
defined as the ratio between ${\rm E_{sc}}$ and ${\rm E_{bb}}$, is
listed in Column 12. 

\subsection{Spectral properties}
\begin{figure}
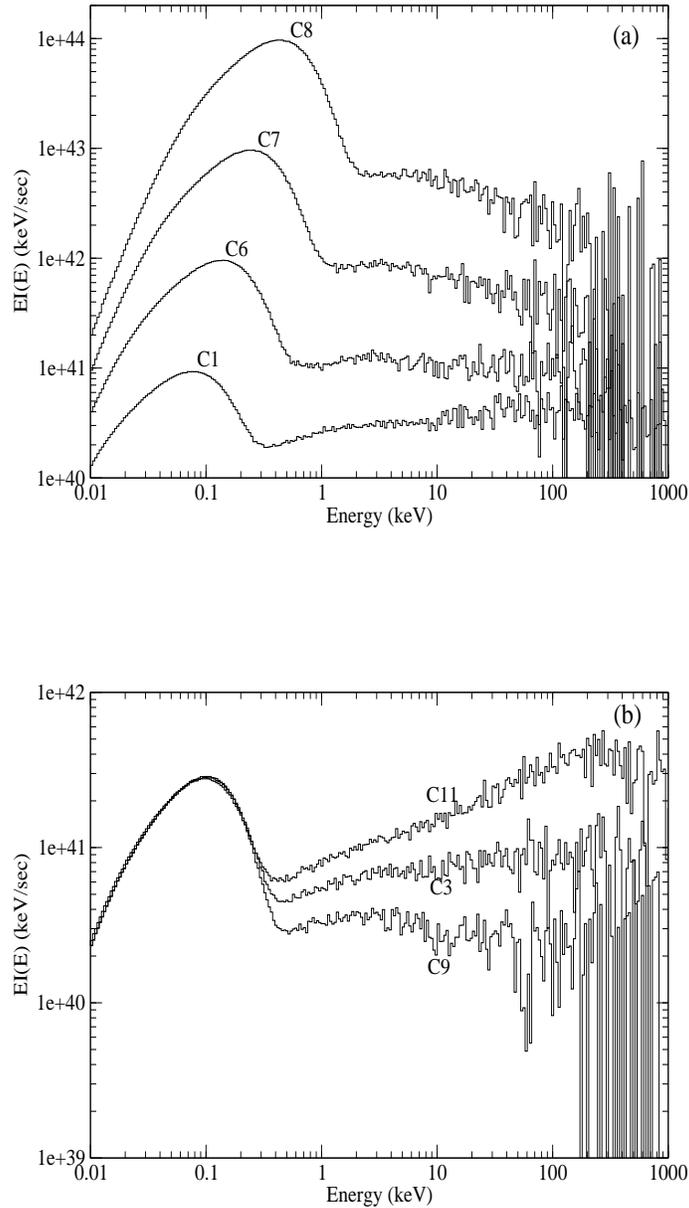

\begin{center}
\includegraphics[width=9cm,height=7cm]{fig2a.eps}
\vskip 2cm
\includegraphics[width=9cm,height=7cm]{fig2b.eps}
\caption{
a) Variation of the shape of the spectrum when 
$\dot{m}_d$ is increased by a factor of 10 starting from $\dot{m}_d = 0.0001$ to $0.1$. 
Case IDs are marked for each plot. The spectrum becomes softer as 
$\dot{m}_d$ is increased.
b) Variation of the spectra when the halo rate $\dot{m}_h$ is increased 
keeping the disk rate constant at $\dot{m}_d = 0.0003$. The spectrum becomes 
harder as $\dot{m}_h$ is increased. 
}
\end{center}
\end{figure}

In Figures 2a and 2b, we show the variation of the shape of the final emergent 
spectra. In Figure 2a, the Keplerian disk rate $\dot{m}_d$ is increased by a factor of 
 10 starting from $\dot{m}_d = 0.0001$, keeping the sub-Keplerian 
halo rate constant at $\dot{m}_h = 0.1$ Eddington rate. In Figure 2b,
the sub-Keplerian rate $\dot{m}_h$ is increased keeping the Keplerian rate 
constant at $\dot{m}_d = 0.0003$ Eddington rate. The case IDs are marked
on each plot. As $\dot{m}_d$ is increased, the relative intensity increases. 
This is understandable since increasing $\dot{m}_d$ increases the number of 
soft photons in a given energy band. At the same time, the number of scattering
among the photons and the electrons increases and hence, the centrifugal 
pressure dominated inner region (including the post-shock 
region, when present) is cooled faster and the region collapses. Thus the volume of the post-shock 
region as well as the number of available hot electrons reduce with the 
increase of $\dot{m}_d$. Since the high energy power-law part of the spectrum 
is determined by the number of available hot electrons as well as the size 
of the post-shock region, the spectra become softer with the increase of $\dot{m}_d$. 
The radiative efficiency of the sub-Keplerian flow is very small compared
to the Keplerian disk as the sub-Keplerian flow is optically thin. Also, 
sub-Keplerian flow has very high radial velocity as compared to the Keplerian flow (Chakrabarti 1989).  
Thus, most of the energy and entropy is advected with the flow itself. 
The Keplerian disk is optically thick and emits black body radiation. 
Thus, the main contribution in the emerging spectrum comes from the Keplerian disk 
and the intensity depends on the Keplerian disk accretion rate ($\dot{m}_d$). 

On the other hand, the spectra become harder as $\dot{m}_h$ is increased. As 
$\dot{m}_h$ is increased, the density of post-shock region increases and 
hence, the optical depth in this region is enhanced. This increases the 
number of scattering and hence, more photons get inverse-Comptonized. So the number of 
high energy photons in a given energy band increase. Thus the spectra become harder. 

For both the cases (increase in either $\dot{m}_d$ or $\dot{m}_h$, 
keeping the other constant), the interaction between the photons and the 
electrons is enhanced and as a result, the electrons get cooler, so the 
shock location moves towards the black hole (see Table 1) as was shown for 
thermal bremsstrahlung earlier (MSC96) and analytically by Das \& Chakrabarti (2004).

\begin{figure}
\begin{center}
\includegraphics[width=10cm,height=8cm]{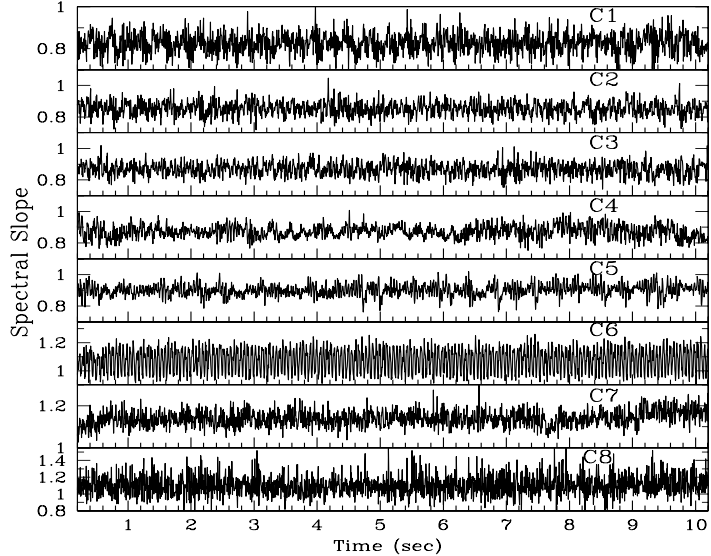}
\caption{Time variation of the slopes of the power-law part of the spectra
is shown. Effects of variations of the disk rate $\dot{m}_d$ is shown. Case
IDs are marked on each panel.
}
\end{center}
\end{figure}

\begin{figure}
\begin{center}
\includegraphics[width=10cm,height=8cm]{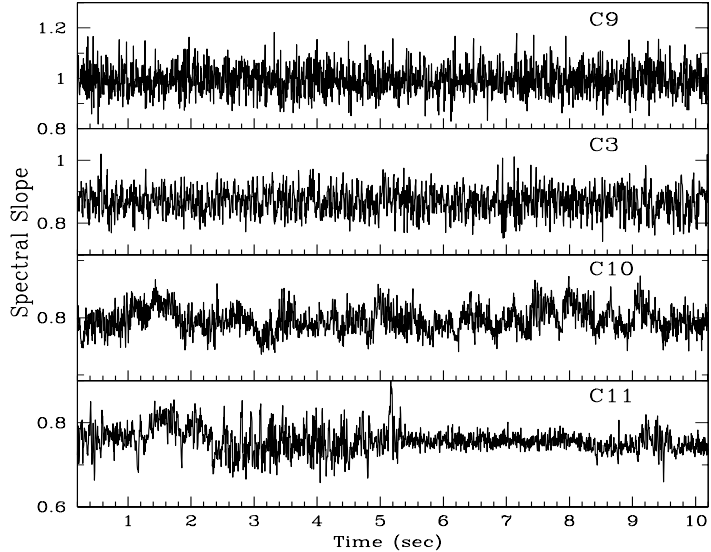}
\caption{Same as Figure 3, but halo rate $\dot{m}_h$ is varied keeping 
$\dot{m}_d$ constant.
}
\end{center}
\end{figure}

In Figures 3 and 4, we show the time variation of the spectral slope $\alpha$
($I(E)\propto E^{-\alpha}$) for all the cases presented in Table 1. The 
case IDs are marked on each panel.  The time averaged value of the spectral 
slopes are given in Table 1. However, when we plot the time variations, we
find an interesting behaviour, namely, the rocking of the spectrum between
hard and soft states. The spectral slope oscillates around or near the value 1.
We find this effect for many cases e.g., C1, C6, C8 and C9. Among these 
cases, we find low frequency QPOs for C6 and C9 (see below).

\subsection{Timing properties}

We compute the time variations of the photon count rates for all the cases in order to generate 
simulated light curves. 
In the coupled simulation run, a Monte-Carlo simulation is run after 
each hydrodynamic time step. The photons are emitted continuously during this time of 
hydrodynamical evolution. However, in our two-step process, we update 
the energy of these photons through the Monte-Carlo simulation step. In the Monte-Carlo simulation, 
we track each photon and as soon as it leaves the disk, it is saved in 
an energy bin as well as in a time bin. We divide the energy range $0.01$ 
keV to $5000$ keV in $300$ logarithmic energy bins. The time resolution 
is of $5$ ms. After the run is completed, we rebin these photons with 
a time bin of $0.01$ s to compute the lightcurves. To count the photons 
in a given energy band, we add up the photon numbers in each energy bins 
which fall in the required energy band. 

\begin{figure}
\begin{center}
\includegraphics[width=10cm,height=10cm]{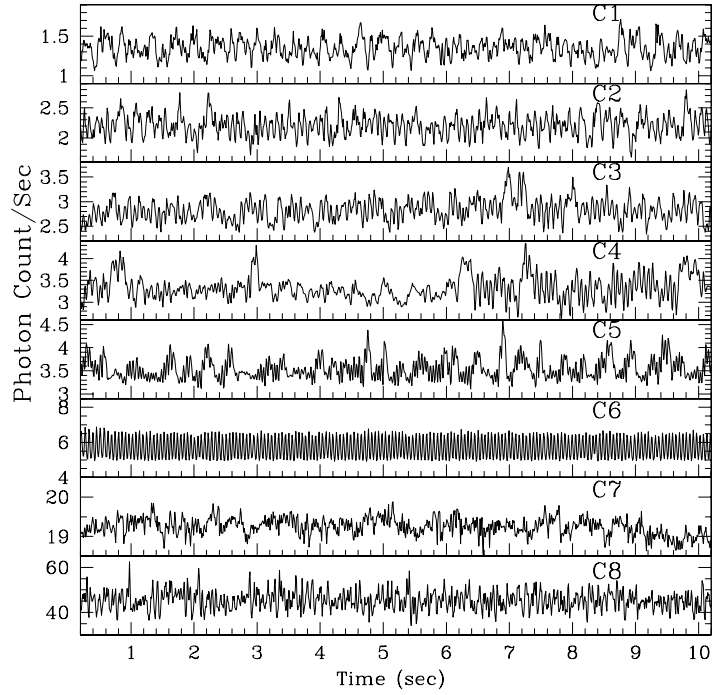}
\caption{The light curves of the photons (in the unit of $10^{42}$) 
which are in the power-law region ($0.5~$keV$~<~E~<100~$keV), are
presented. Here, $\dot{m}_d$ is increased keeping $\dot{m}_h=0.1$ constant. 
See text for the detailed computational procedure of the light curves.
}
\end{center}
\end{figure}

\begin{figure}
\begin{center}
\includegraphics[width=10cm,height=10cm]{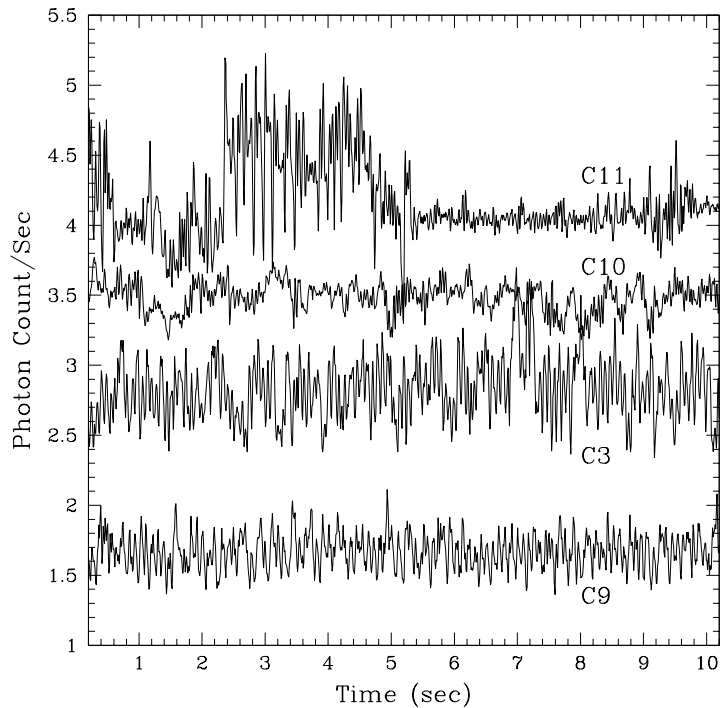}
\caption{Same as Figure 5, but $\dot{m}_h$ is increased keeping 
$\dot{m}_d$ constant at $0.0003$ Eddington rate.}
\end{center}
\end{figure}

In Figures 5 and 6, we plot light curves of the photons in the energy 
band 0.5 keV to 100 keV (for C7 and C8, 2 keV$~<~E~<~100$ keV). The photons in 
this energy range are mostly the inverse-Comptonized photons. The case IDs 
are marked on each plot. In Figure 5, the Keplerian rate $\dot{m}_d$ is 
increased keeping the sub-Keplerian rate $\dot{m}_h$ constant, whereas 
in Figure 6, $\dot{m}_h$ is increased keeping $\dot{m}_d$ constant. 
In both the Figures, we see that the count rate increases with the increase 
of the variable parameter (e.g. $\dot{m}_d$ or $\dot{m}_h$). For Figure 5, 
its understandable since increasing $\dot{m}_d$ increases the number of 
available soft photons. On the other hand, when we increase $\dot{m}_h$, 
the optical depth of the post-shock region increases and hence, the interception 
of the photons by the electrons increases. Thus the number of Comptonized 
photons increases and that explains the increase of count rates with $\dot{m}_h$
in Figure 6.  The variations in the lightcurves are arising because of 
the variations in the hydrodynamic and thermal properties of the post-shock region. 

\begin{figure}
\begin{center}
\includegraphics[width=10cm,height=10cm]{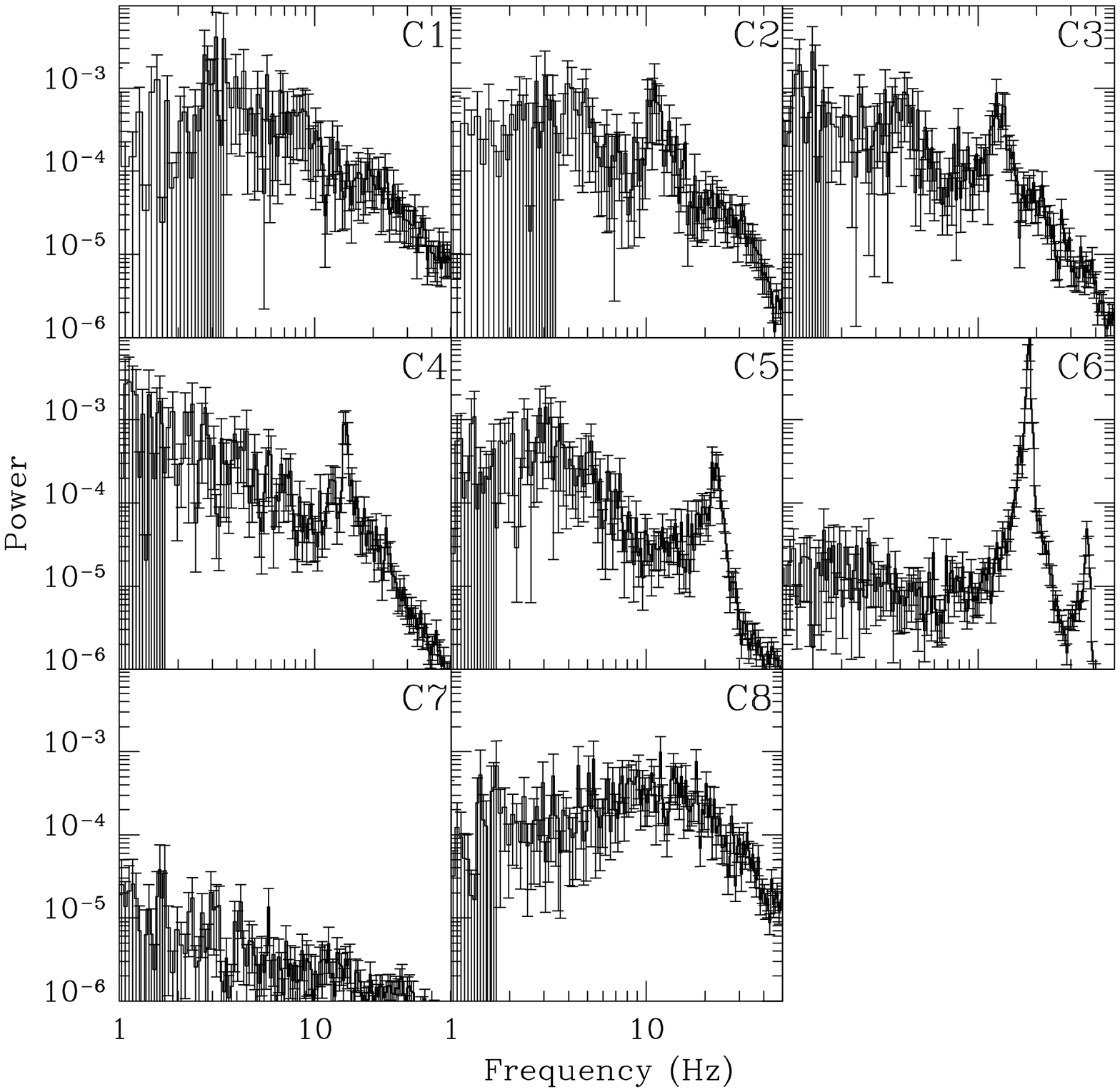}
\caption{Power Density Spectra (PDS) of the all cases presented in 
Figure 5. QPO frequency increases with the increase of $\dot{m}_d$.}
\end{center}
\end{figure}

\begin{figure}
\begin{center}
\includegraphics[width=10cm,height=10cm]{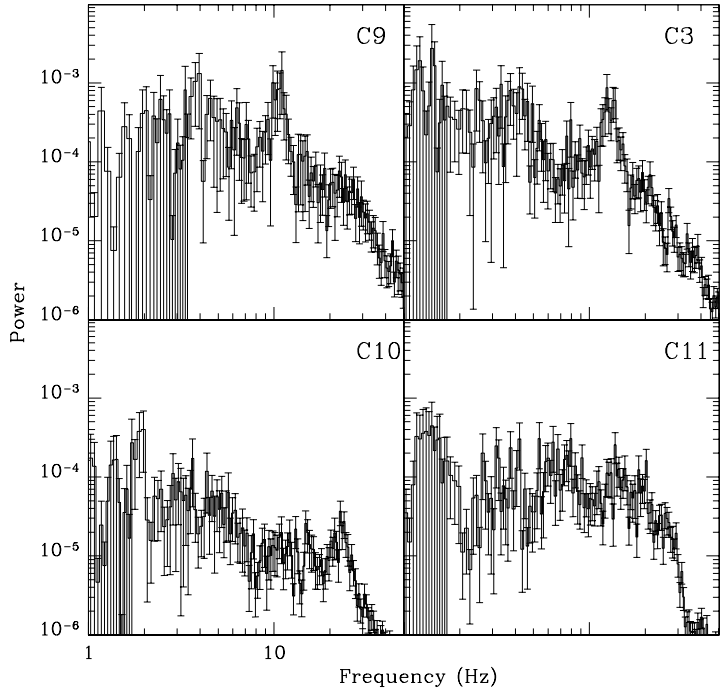}
\caption{Power Density Spectra (PDS) of the all cases presented in 
Figure 6. QPO frequency increases with the increase of $\dot{m}_h$.}
\end{center}
\end{figure}

In Figures 7 and 8, we show the Power Density Spectra (PDS) for all the cases. 
The case IDs are marked in each panel. We find low frequency quasi 
periodic oscillations (LFQPO) for some cases. The frequencies are listed 
in Table 1. We find that LFQPO frequencies increase with the increase of 
both $\dot{m}_d$ and $\dot{m}_h$.  
We have seen that the spectra become 
softer with the increase of $\dot{m}_d$ and with the decrease of $\dot{m}_h$.
Therefore from Figures 3 and 7, we find that the LFQPO increases as the object transits 
from the harder state to the softer states. We find the opposite behavior 
in Figures 4 and 8.  The trend of the enhancement factor (Q) 
in Table 1 is exactly the opposite of the trend in QPO frequency, which is reasonable. 

It has been argued in the literature (MSC96; CM00; CAM04; Debnath et al. 2010) that the 
oscillation in the centrifugal barrier dominated hot region is responsible for the LFQPO observed in the 
black hole candidates. In this model, LFQPO arises when the infall time scale of post-shock 
matter roughly matches with the cooling time scale. For the present simulations, 
we find the shape of this hot post-shock region to be mostly  paraboloid with the base 
to be oblate spheroid sometimes. Turbulence and backflow of matter are 
present sometimes in the post-shock region. Therefore, it is not always 
easy to calculate the exact infall time scale. In our grid based hydrodynamic 
simulation, we compute the infall time scale in the following way. We take 
the average of the radial velocity component over 20 vertical grids starting 
from the equatorial plane at each radius and compute a radial velocity 
profile near the equatorial plane. 
$$v_{avg}(R)\sim\frac{\sum_{iz=1}^{20} v_R(R,iz)}{20},$$
where, $v_R(R,1)$ represents the radial component of the velocity on the
first grid (i.e. on the equatorial plane), $v_R(R,2)$ represents the 
same on the second grid and so on. Then we calculate the infall time scale as 
$$t_{in} =\sum_{R=R_{sh}}^{1.5}\frac{dR}{v_{avg}(R)}.$$
The cooling timescale $t_{cool}$ is easy to compute as it includes only 
the scalar quantities. 
$$t_{cool}=\frac{E_{th}}{\dot{E}}.$$ 
Here, $E_{th}$ is the total thermal energy in the post-shock region and 
$\dot{E}$ is the cooling rate in the same region, which we calculate directly
from the Monte-Carlo simulation. In Table 1, we present the ratio 
$t_{in}\over t_{cool}$ in the last column. We see that this ratio is nearly 
1 for all the cases when LFQPOs are seen. Thus the proposal of LFQPOs arising out 
of resonance oscillation (MSC96, CAM04) of the post-shock region appears to be justified.
However, earlier, only the power-law cooling was used as the proxy to the Compton cooling.
In the present paper, actual Comptonization has been used. 
    
\section{Concluding Remarks}

There were several conjectures in the literature regarding the spectral and timing properties 
of a black hole, which appeared to be plausible, 
but were not proven rigorously with time dependent codes. These were: State transitions are possible 
by variations of the disk and halo accretion rates in a two-component advective 
flow (TCAF) solutions (CT95); Low Frequency quasi-periodic oscillations 
(LFQPOs) were the results of oscillation of centrifugal pressure dominated shocks 
in a transonic flow and the oscillations are due to resonance between the 
cooling and infall time scales (MSC96; CAM04). In a series of papers we 
have shown the effects of radiative transfer (through Monte-Carlo method) 
on the hydrodynamic solution (Ghosh et al. 2011; Garain et al. 2012). In the 
present paper, we furthered our earlier results to show that indeed the spectral 
and timing properties of a TCAF are exactly as conjectured before. Specifically, 
after running a large number of model cases with various combinations of the 
accretion rates, we have been able to show that QPOs are observed when oscillations 
are induced into the flow by resonance. 

In the present paper, we have not introduced the viscosity in the flow. A high 
viscosity in the equatorial plane produces a Keplerian disk (Giri \& Chakrabarti 2013) 
automatically while lower viscosity away from the equatorial plane fails 
to convert the sub-Keplerian 
flow into a Keplerian disk. This was predicted long ago (Chakrabarti 1990). 
Instead of adding viscosity to the flow, we directly included a Keplerian 
disk as the supplier of the seed photons in the equatorial plane. The most self-consistent 
solution needs to incorporate the viscosity and produce the Keplerian disk, 
and its spectrum {\it ab initio}. This requires generation of photons through 
bremsstrahlung process and Comptonize them to produce the multicolor black 
body spectrum from the Keplerian disk. Presently, the progress is limited 
by computational constraints. 

Some of the hard X-rays produced in the Compton cloud will be intercepted back by the
cold disk. These hard photons will be absorbed by the disk and may increase 
its temperature. This effect, in turn, will produce some more soft photons in addition 
to the number of photons computed assuming SS73 prescription, which may give additional 
cooling to the sub-Keplerian flow and soften the spectra. In the context of steady solutions, CT95 incorporated these
so-called reflection effects self-consistently. In the present context of time dependent solutions too, we have seen in our 
simulation that approximately 50\% of the photons which suffered scattering inside the 
Compton cloud are getting absorbed by the Keplerian disk and are re-emitted at a lower energy. We also find that 
the spectrum is becoming mildly softer due to this effect but not enough
to influence the final conclusions of this paper. 

\section{Acknowledgment}

The work of HG was supported by a post doctoral grant from Ministry of 
Earth Science, Government of India.
{}
\end{document}